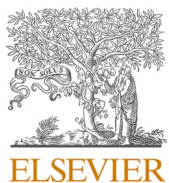
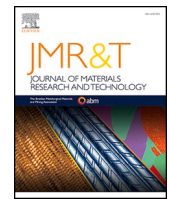

# Tailoring high-entropy alloys via commodity powders for metal injection moulding: A feasibility study

A. Meza [a,b,*], A. Barbosa [b], E. Tabares [b,c], J.M. Torralba [a,b]

[a] *IMDEA Materials Institute, Eric Candel 2, 28906, Getafe, Spain*
[b] *University Carlos III of Madrid, Av. Universidad 30, 28911, Leganés, Madrid, Spain*
[c] *Tekniker, Basque Research and Technology Alliance (BRTA), C/ Iñaki Goenaga, 5, 20600, Eibar, Spain*

A B S T R A C T

High entropy alloys (HEAs) represent a novel frontier in metallurgical advancements, offering exceptional mechanical properties owing to their unique multicomponent nature. This study explores a novel strategy utilising commodity powders - Ni625, Invar36, and CoCrF75 - to tailor HEAs via metal injection moulding (MIM). The objective is to achieve cost-effective manufacturing while maintaining desired properties. The research involves mixing these powders in a specific proportion, integrating them with a sustainable polyethylene glycol-cellulose acetate butyrate binder system, and characterising the resulting feedstocks for MIM processing. Subsequent debinding and sintering steps were executed to densify and form a single face-centered cubic (FCC) phase HEA, followed by comprehensive analyses to evaluate the suitability of the developed HEA compositions. In addition, all MIM stages were thoroughly characterised to control the porosity of the final parts and to ensure a single FCC solid solution with promising mechanical properties in the developed non-equiatomic CoCrFeNiMo$_x$-type HEAs.

## 1. Introduction

High entropy alloys (HEAs) have attracted considerable attention in recent years due to their exceptional mechanical attributes. The pioneering works by Cantor et al. [1] and Yeh et al. [2] in 2004 introduced the concept of HEAs, where five or more elements are blended in relatively high concentrations (5–35 at. %); being one of the important features that it crystallises in a single phase, often body-centered cubic (BCC) or face-centered cubic (FCC), or mix of both, driven by the high entropy of mixing [3–6]. These alloys are opening up uncharted territory in the field of materials engineering, offering appealing properties such as high strength at high temperatures and increased ductility [7].

The development of HEAs through powder metallurgy routes presents several advantages over casting processes by avoiding elemental segregation, refining grain size, and accommodating metals with dissimilar densities or extremely high melting points [8,9]. Typically, producing HEAs via powder metallurgy involves using either pure elemental powders or prealloyed powders with the specific HEA composition. However, certain prealloyed powders derived from commercial alloys like Ni625, Invar36, or CoCrF75 already contain elements conducive to HEA formation. Produced on a large scale, these powders are more cost-effective than highly pure or customised prealloyed powders. Hence, this work introduces a novel strategy that selects different commercial or "commodity" powders to tailor the final composition of manufactured HEAs [10].

Much of the research has been focused on the development and characterisation of various HEAs to explore their potential applications. Metal injection moulding (MIM) has emerged as an alternative manufacturing technique for producing components with desired final geometries. MIM allows the production of intricately shaped samples with high reproducibility at a reduced cost [11]. In preparing HEA powders for MIM, attention is given to adapting the powder production to the injection process, as the initial powder characteristics (such as particle size and morphology) have a significant impact on MIM processing [12]. Additionally, choosing a binder system plays a crucial role in providing the feedstock with requisite viscosity properties for proper mould filling. Moreover, this polymer system must be removed from the samples before sintering to preserve structural integrity while preventing defect formations. This study focuses on characterising and tailoring commodity powders for MIM processing. It involved producing various feedstocks by combining the selected powders in different loadings with a sustainable binder comprising polyethylene glycol (PEG) and cellulose acetate butyrate (CAB) while rigorously characterising and controlling all steps of the processing route. This binder system was chosen for its eco-friendly nature [13,14]. The PEG polymer is water-soluble, minimising environmental impact during the solvent-debinding process. Moreover, the CAB polymer is sourced from natural polymers, offering a sustainable alternative to conventional synthetic polymers [15,16]. The resulting green samples underwent a two-step debinding procedure. Debinding is the process of removing the binder from the green parts, leaving behind a porous metallic structure. Two-step debinding is a common method to prevent the formation of cracks in the metallic parts during sintering. In the first step, the green parts are immersed in a solvent to remove the water-soluble PEG binder. In the second step, the green parts are heated to a temperature where the CAB binder decomposes. Finally, once the polymers were successfully removed, the samples were sintered and densified, forming a single FCC microstructure, which provided excellent properties after microstructural and mechanical characterisations.

## 2. Experimental procedures

Various ratios of gas-atomised commodity powders, including Ni625 (VDM Metals), Invar36 (Sandvik Osprey), and CoCrF75 (Mimete), were combined to attain the desired final composition for the high entropy alloy (MIM_HEA), as detailed in Table 1. The calculation of the slope parameter ($S_w$) was conducted to ensure that the amalgamation of these three commodity powders resulted in a broad particle size distribution (particle size distribution examined with a Bettersizer ST Laser Particle Size Analyser). Such broad distribution is advantageous as it contributes to achieving reduced viscosity when incorporated with the binder system [17]. A broader particle size distribution aligns with ideal conditions for the MIM process, with $S_w$ values ideally ranging from 2 to 4, thereby enhancing the alloy's suitability for injection moulding.

Once mixed, the powders were incorporated into the binder system at various loadings using a Haake Polylab mixer (Thermo Scientific). The binder formulation, relying on PEG and CAB components, had undergone prior optimization and validation in earlier studies [15].

Afterwards, the rheological properties of the prepared feedstocks were examined under similar temperatures to the injection process, at 180 °C, via torque rheology, viscosity measurements, and melt flow index evaluations. A preliminary test using a Haake Polylab QC mixer (Thermofisher) provided values for the critical solid loading around 70 vol% loading. Thus, several solid loadings were chosen around this critical solid loading, and torque rheology was investigated while maintaining a consistent volume for all the feedstocks to select the feedstock with the highest powder loading that still maintained a stable torque over time. The rollers were adjusted to 50 rpm, and torque stabilization was observed for a duration of 30 min. Viscosity measurements were conducted using Haake MiniLab 3 equipment (Thermo Scientific), equipped with a pressure sensor capable of assessing the feedstock viscosity relative to applied shear rates. The temperature was fixed at 180 °C, and shear rates were adjusted by increasing the rotational speed of the roller screws from 25 to 250 rpm in increments of 25 units for each measurement. To ensure result reliability, three measurements were taken for each feedstock. Finally, prior to the injection phase, the flowability of the developed feedstocks was assessed through melt flow index (MFI) measurements using an MFI tester, Kejian 3092 (Kinsgeo). At 180 °C and applying a 1.5 kg weight, the mass of feedstock able to flow within 10 min was measured. For successful injection, the flow value needed to exceed 20 g/10 min.

Following the injection of the feedstocks into a tensile test-shaped mould, using the Mini-injector Xplore IM12 (Xplore) at 180 °C and a pressure of 6 bars, the produced green samples underwent a two-step debinding process. Initially, the samples were immersed in agitated water at 60 °C for 1 h to eliminate the PEG component. Subsequently, after drying, thermal debinding occurred in a specialised furnace (GD-DC-50, Goceram) under a protective argon atmosphere, gradually heating to 500 °C. The heating and cooling rates were set at 0.5 and 1 °C/min, respectively, so that the CAB degradation took place gradually, preventing the formation of cracks caused by sudden gas expansion. To investigate the degradation temperature range of the binder system, thermogravimetric analyses were conducted using a Simultaneous Thermal Analyzer (STA) STA6000 (PerkinElmer). A heating rate of 10 °C/min was applied under an Ar atmosphere until reaching 850 °C. These analyses were performed on green, solvent-debound, and brown samples derived from each optimized feedstock.

Following the complete removal of the binder, the sintering process occurred in a Vacuum Furnace (Nabertherm) operating under an Ar atmosphere at 1200 °C for a sintering time of 6 h. The selection of these parameters was based in previous works in which an optimization of the annealing heat treatment for the homogenisation of this HEA composition was developed [18]. After sintering, the relative density of the samples was studied using the Archimedes method.

The samples underwent meticulous characterisation throughout various stages of the MIM process. This involved microstructural analysis via scanning electron microscopy (SEM, Helios Nanolab 600i), including energy-dispersive X-ray spectroscopy (EDS) to check the elemental distribution, electron backscatter diffraction explorations (EBSD) to analyse the grain microstructure, and the utilisation of X-ray diffraction (XRD, PANalytical) to confirm the formation of a single FCC solid solution encompassing all constituent elements.

Finally, the sintered samples were subjected to mechanical characterization through hardness measurements using an HVM-2T E instrument (Shimizu). A 500 gf applied load was employed for 15 s, conducting 10 measurements on each sample. Furthermore, tensile tests were tested on the densified samples at room temperature (RT) using an Instron 3384 universal electromechanical testing machine. A 10 kN load cell was utilized, and the tensile test speed was set at 0.015 mm/s. A total of 10 samples were tested for each selected feedstock. Finally, fractography images were taken with SEM to evaluate the brittle or ductile fracture mode of the tensile-tested specimens.

## 3. Results and discussion

Fig. 1 shows the rheological properties of the feedstocks produced. The torque rheology of powders with binders, as depicted in the left side of Fig. 1, facilitated a continuous assessment of the mixture's stability over time. It is evident that all feedstocks consistently exhibited low torque values, confirming their flowability at 180 °C and 50 rpm, even with high concentrations of metallic powders (up to 77 vol%). The stabilization of torque indicated successful blending and uniform dispersion of metallic powders within the chosen binders. However, at a 77 %vol., torque values exhibited a gradual increase, suggesting that the powder fraction surpassed the binding polymers' capacity for thorough blending, resulting in the exclusion of this specific feedstock from subsequent analysis.

**Table 1**
Commodity powders and MIM_HEA features.

| Alloy | Powder % | wt. (%) | | | | | | $d_{10}$ (μm) | $d_{90}$ (μm) | $S_w$ |
|---|---|---|---|---|---|---|---|---|---|---|
| | | Ni | Fe | Cr | Mo | Co | Nb | | | |
| Ni625 | 20 | 61.5 | 5.4 | 25.3 | 5.6 | – | 3.8 | 21.9 | 47.8 | 7.57 |
| Invar36 | 38 | 34.8 | 65.2 | – | – | – | – | 2.3 | 6.4 | 5.68 |
| CoCrF75 | 42 | 0.5 | 0.8 | 32.6 | 3.7 | 62.4 | – | 20.5 | 46.3 | 7.24 |
| at. (%) | | | | | | | | | | |
| MIM_HEA | – | 25.7 | 26.6 | 19.3 | 3 | 25 | 0.4 | 2.17 | 48.8 | 1.89 |

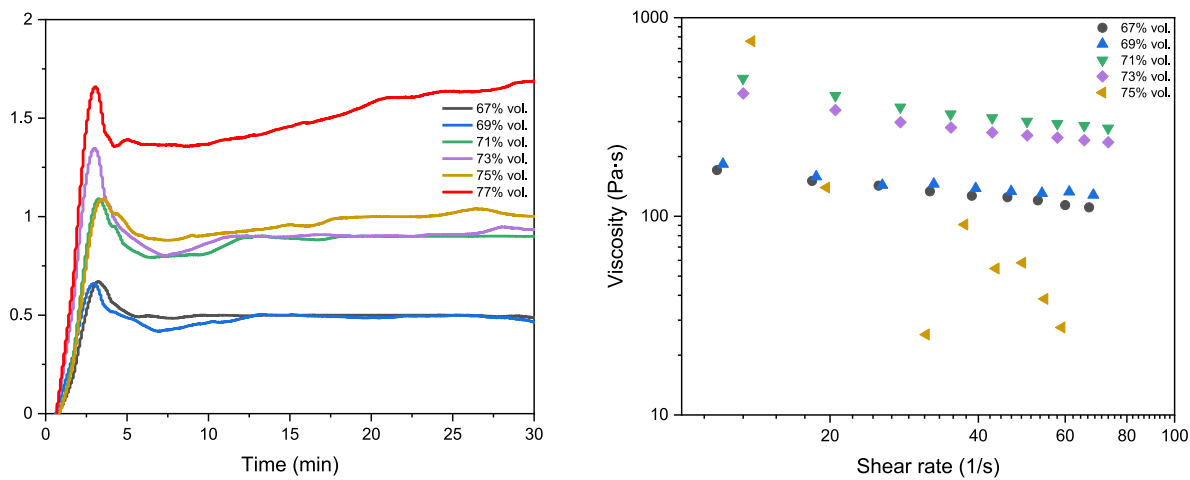

**Fig. 1.** Rheology of the feedstocks measured by Torque evolution during mixing (left) and viscosity measurements measured in an extrusion process (right).

Moreover, the viscosity of the feedstocks was measured under increasing shear rates at 180 °C (as depicted on the right side of Fig. 1). This was done to identify any pseudoplastic behaviour demonstrated by the feedstocks, characterized by decreasing viscosity as the shear rate increased. Post-analysis revealed that all feedstocks displayed this pseudoplastic behaviour. Furthermore, viscosity values consistently augmented in proportion to higher powder loadings. Nonetheless, these viscosity values consistently remained below 1000 Pa s, aligning with the ideal range conducive to an efficient injection process [11].

Furthermore, MFI measurements were conducted on the 71, 73 and 75 vol% loaded feedstocks (Fig. 2), selected as having the highest metallic loading capable of ensuring increased density in the final sintered samples while maintaining suitable rheological properties for injection. These MFI values go in concordance with the rheological properties of the mixtures. With higher viscosity, the amount of material capable of being poured into the mould decreases.

Consequently, based on these findings, the feedstocks containing metallic volume fractions of 71, 73, and 75 vol% were chosen for the injection process. SEM images depicting the cross-sections of injected green samples are presented in Fig. 3. A thorough examination of all injected samples revealed an effective dispersion of powder particles within the binders. The smaller Invar36 powder particles adeptly filled the spaces between the larger Ni625 and CoCrF75 particles, facilitating a convenient homogenisation of the powders and preventing agglomerate formation. These characteristics contribute significantly to achieving reduced porosity levels post-sintering of the debound samples while ensuring the correct synthesis of the HEA.

The STAs conducted on the injected samples in their green state, blue state (after solvent debinding), and brown state, as depicted in Fig. 4, lead to the conclusion that an optimal debinding temperature range falls between 440 and 450 °C. Within this temperature range, all organic masses have undergone decomposition, and beyond this point, there are no significant abrupt mass losses observed with increasing temperature. In the straight lines representing the green samples for all feedstocks, two distinct slopes can be identified during the degradation process. This divergent behaviour is attributed to the different degradation behaviours of the two polymers (PEG + CAB) at this stage of the MIM process. Notably, these two slopes are absent in the blue samples (post-solvent debinding), signifying the effective removal of the PEG.

Moreover, the STAs conducted on the blue samples reveal a notable decrease in mass loss compared to the green samples, suggesting the efficient removal of the CAB polymer during solvent debinding. A comparable trend is observed in the STAs of the brown samples, showcasing nearly negligible mass loss. This minimal mass loss signifies the complete elimination of polymers following the thermal debinding process.

Fig. 5 shows the SEM images taken from the final sintered MIM_HEA feedstocks with 71–75 vol% of powder loading. It is clear how, by having a higher powder fraction, improved densification levels were achieved (up to 99.2 % relative density for the 75 %vol. MIM_HEA_75%).

Furthermore, the compositional mappings demonstrate a proper homogenisation of all the elements (Fig. 6). This homogeneous distribution indicates that a single phase has been obtained just after the sintering of the brown parts without the need for additional homogenisation or annealing heat treatments. This homogenisation was observed in all the sintered samples, no matter the feedstock used. Nevertheless, some minor Cr carbides were detected in the grain boundaries.

Nevertheless, to certify that a single phase was obtained after the sintering, XRD analysis was carried out (Fig. 7). From these results, a single FCC microstructure was detected in the sintered samples. Thus, it can be concluded that after the MIM process, it was possible to attain HEAs using mixes of commodity powders.

The EBSD explorations also showed the achievement of a single FCC phase in the sintered samples (Fig. 8). Th < ese mappings also confirmed that the equiaxed grain sizes of these samples were centered at 19 ± 7.5 μm, which is in accordance with other studies from the literature [19]. Also, no texture was detected in the sintered samples. The appearance of small grain colonies surrounded by areas of bigger grains can be due to the surrounding of some of the bigger CoCr75 powder particles by the smaller Invar36 particles. In turn, the grains of these coarser particles didn't grow as much. However, this phenomenon did not impede the diffusion of elements, facilitating the formation of a single FCC phase HEA.

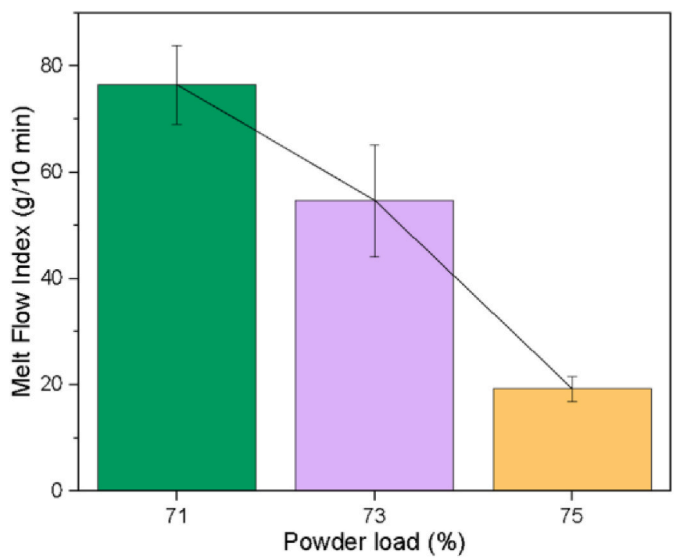

**Fig. 2.** MFI measurements of the selected feedstocks at 180 °C.

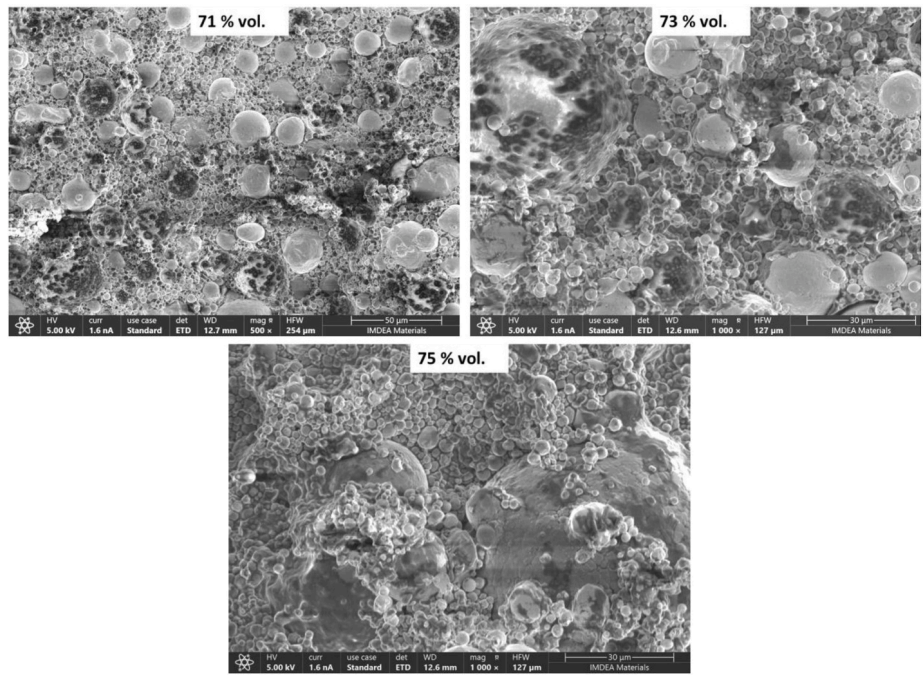

Fig. 3. SEM Images of the cross-section of green injected parts.

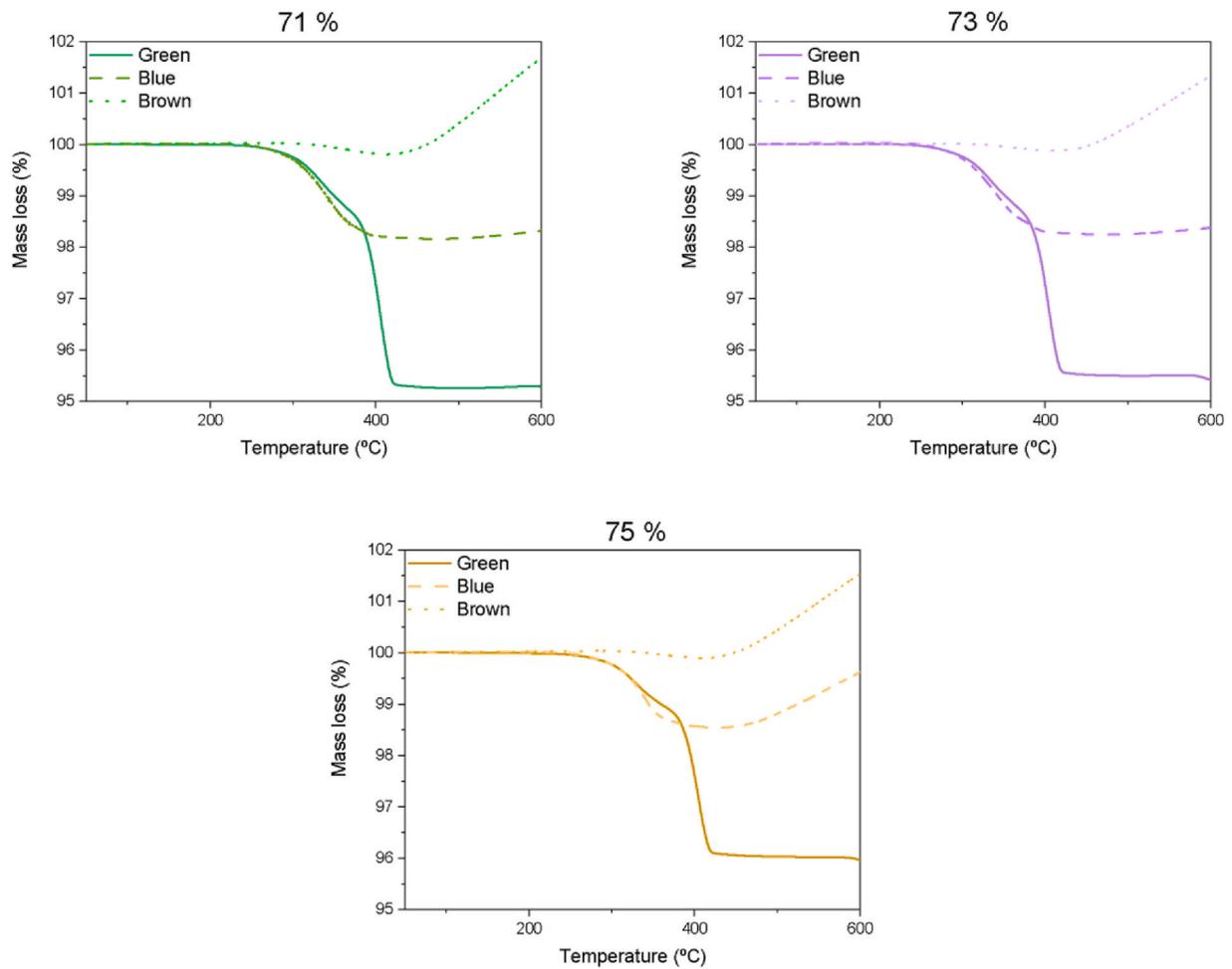

Fig. 4. STAs of the green, blue and brown samples.

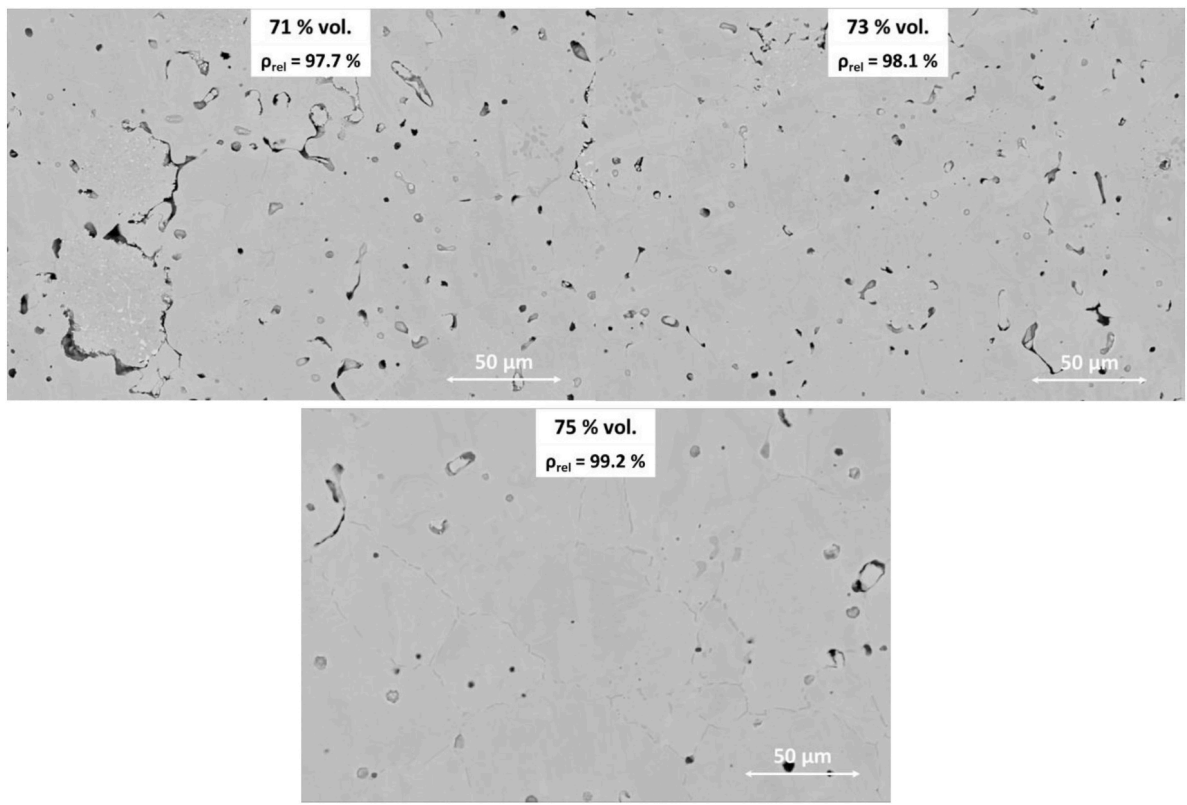

**Fig. 5.** SEM Images of the sintered parts at 1200 °C for 6 h.

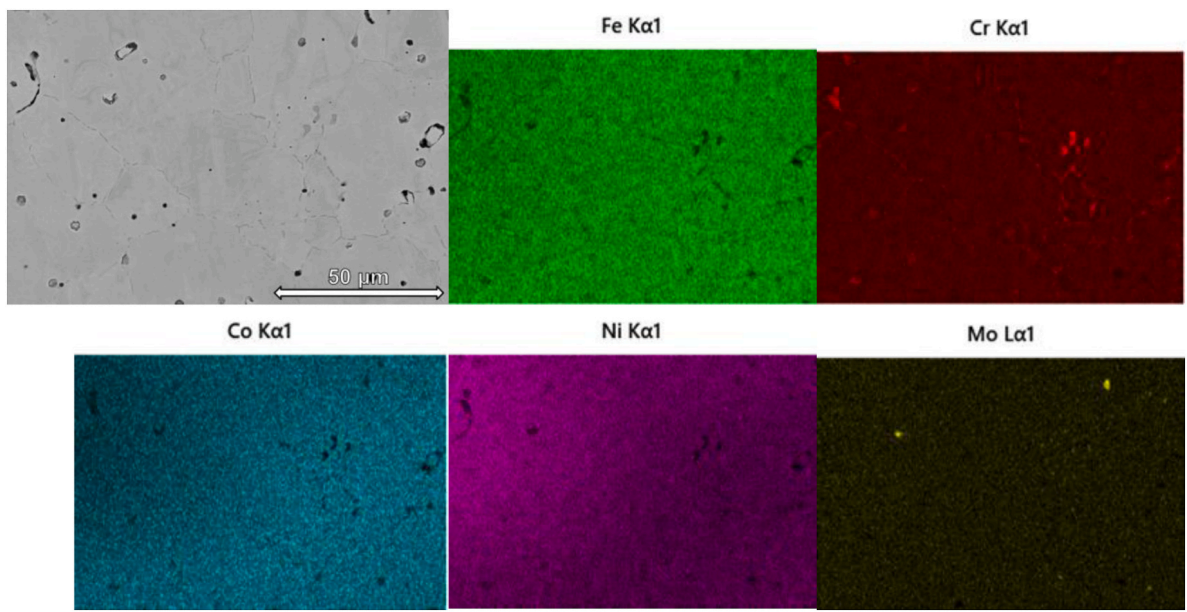

**Fig. 6.** SEM/EDS mappings of the MIM_HEA_75% loading after sintering.

The tensile response at RT of the studied materials is plotted in Fig. 9, while their average mechanical properties are summarized in Table 2. Upon initial observation, the samples demonstrate a notable ratio between ultimate tensile strength (UTS) and ductility, particularly noticeable in the MIM_HEA_75% samples, showing superior densification owing to their higher powder loading. This reduction in the porosity involved an improvement of the overall mechanical performance of the final components, as the formed cracks will have fewer points to propagate along the grain microstructure. Remarkably, these samples exhibited outstanding values for both UTS and ductility among MIM-processed HEAs, particularly when compared to the only existing study to date involving a HEA Cantor alloy - very similar in composition to the HEA in this study - processed by MIM (Yield stress = 225 MPa, Ultimate tensile strength = 421 MPa, ductility = 12.6%) [19]. The presence of Mo in the composition of this work's HEA is responsible for achieving improved UTS and ductility because this element further distorts the FCC microstructure, contributing to the lattice distortion strengthening. The mechanical performance is enhanced through this mechanism by impeding the movement of dislocations through stress fields present in the distorted lattice. This mechanism is particularly significant in HEAs, given the nearly equiatomic proportions of elements and different element sizes, especially when Mo is present [20].

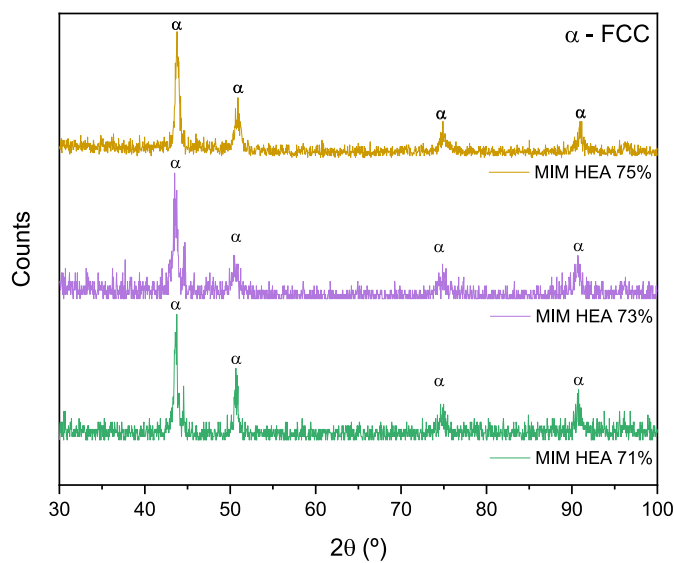

**Fig. 7.** XRD of the sintered parts injected with different metallic loadings.

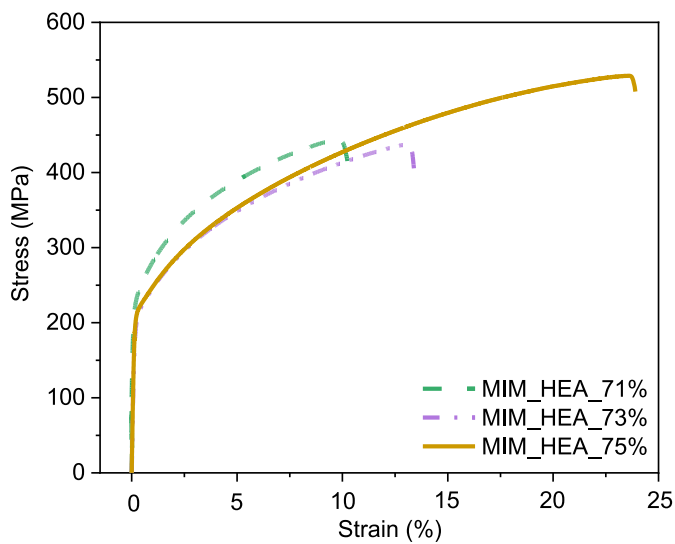

**Fig. 9.** Stress-strain curves obtained from the tensile tests at room temperature.

**Table 2**
Yield strength, Ultimate tensile strength, Ductility and Hardness of the samples tested at RT.

| Sample | Yield Strength (MPa) | Ultimate Tensile Strength (MPa) | Ductility % | Hardness $HV_{1Kg}$ |
| --- | --- | --- | --- | --- |
| MIM_HEA_71% | 186 ± 30 | 330 ± 113 | 5.7 ± 4.5 | 182.7 ± 12.2 |
| MIM_HEA_73% | 196 ± 10 | 459 ± 36 | 10.9 ± 2.7 | 183.1 ± 10.8 |
| MIM_HEA_75% | 204 ± 10 | 478 ± 50 | 18.6 ± 5.5 | 188.8 ± 14.6 |

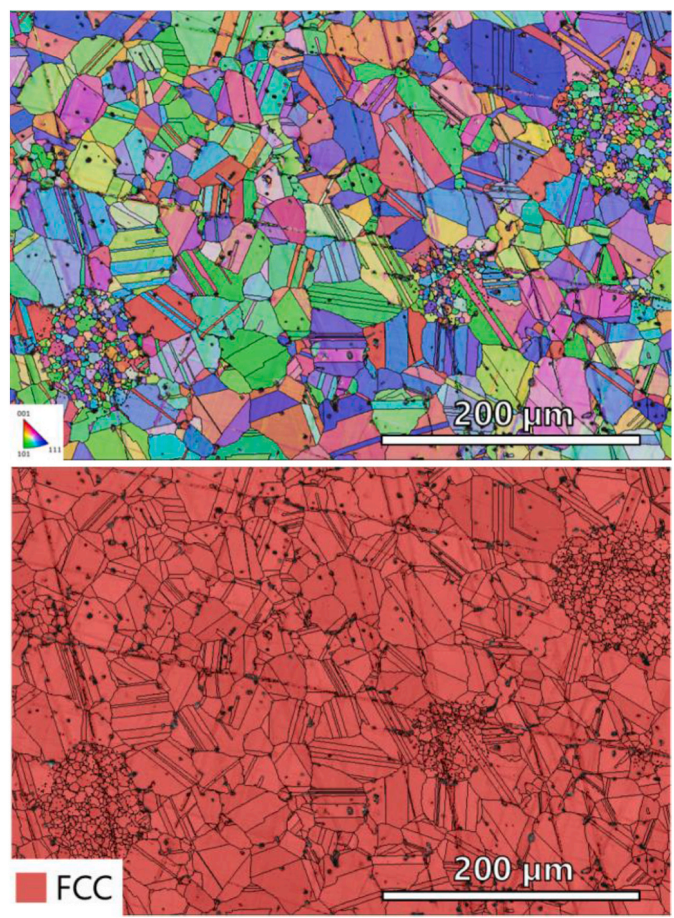

**Fig. 8.** Inverse Pole Figure in X axis (IPF X, up) and phases map (down) of the MIM_HEA_75%.

Thus, these results prove to be very promising and open up the potential to developing high-performance HEAs by MIM.

Nevertheless, we have found a large dispersion in the obtained results. After thorough investigation, this high variability was associated with the formation of debinding cracks in some of the samples coming from the solvent debinding step. These cracks are detrimental to the mechanical performance of the sintered samples, as observed in the fractography of the tested specimens, and must be avoided through better debinding optimization.

In addition, the hardness of the sintered samples fabricated from the feedstocks with different powder loadings was assessed (Table 2). As expected, higher densification of the samples conveyed enhanced hardness of the sintered samples due to a lower porosity in them.

Fig. 10 displays examinations of the fracture surface of the tested samples. The presence of debinding cracks, as mentioned earlier, led to a brittle fracture observed in the macroscale, exemplified by the fracture images of MIM_HEA_71%. Consequently, these cracks hindered the samples from achieving their maximum mechanical properties. However, upon closer examination at a higher magnification on the microscale of the fracture surface, numerous dimples become evident. These dimples serve as clear indicators of high plastic deformation before fracture, revealing the ductile behaviour of the samples. Thus, in the case of improving the solvent debinding step, enhanced and more consistent mechanical performance should be expected from these MIM-manufactured HEAs.

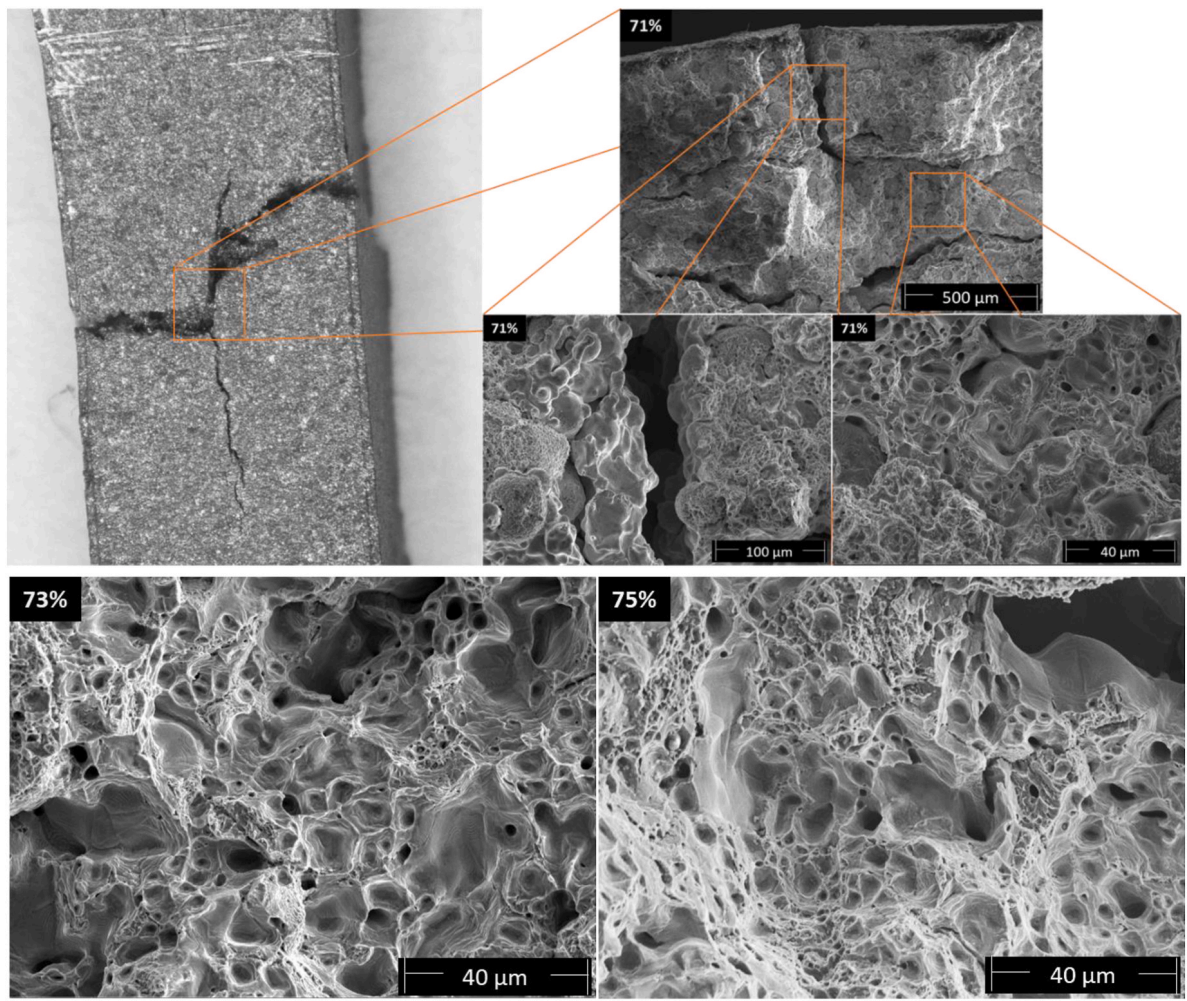

**Fig. 10.** SEM Images of the fracture surface of the tested samples.

## 4. Summary and conclusions

Commodity powders, including Ni625, Invar36, and CoCrF75, were successfully employed to develop a CoCrFeNiMo$_x$ high entropy alloy (HEA) using a metal injection moulding (MIM) route with a polyethylene glycol-cellulose acetate butyrate (PEG/CAB) binder system. After injection, debinding and sintering, a single face-centered cubic (FCC) phase with proper mechanical properties has been obtained. The results indicate that:

- ➢ The powders characteristics facilitated their incorporation into the PEG/CAB binders in high loadings (71, 73, and 75 %vol.) thanks to their broad particle size distribution. Achieving high loadings involved a higher densification of the samples during the sintering stage and better mechanical performance.
- ➢ The sintering process at 1200 °C for 6 h effectively densified the samples and homogenised the elements, forming an FCC HEA in a single step.
- ➢ MIM-processed parts exhibited an exceptional strength-to-ductility ratio with high ductility values. Although further optimization of the debinding stage should be performed to potentially enhance the overall mechanical behaviour of the fabricated HEA parts, this work sets a novel approach to obtain near-net-shaped HEA parts starting from commodity powders.

- ➢ This work is a feasibility study that demonstrates the viability of the proposal. Further studies must be done to optimize the process, specially the solvent and thermal debinding steps, to ensure higher reproducibility.

## Declaration of competing interest

The authors declare that they have no known competing financial interests or personal relationships that could have appeared to influence the work reported in this paper.

## Acknowledgements

The research leading to these results has received funding from the coordinated project "High Entropy Alloys Resistant to Hydrogen Embrittlement" (EARTH) (ref. TED2021-130255B–C31), funded by MCIN/AEI/10.13039/501100011033 and by the European Union "NextGenerationEU"/PRTR. Diego Iriarte, Xiaomei Yang, Jimena de la Vega, and Amalia San Román are sincerely thanked for their help in this research.